\documentclass[12pt]{iopart}
\newcommand{\Sch}{{Schr$\ddot{\rm o}$dinger equation }}
\begin{document}

\hfill{\normalsize\sf FIAN/TD/02-13}\\

\vspace{1cm}
\title{Singular Centre in Quantum Mechanics
 as a Black Hole}

\author{A E Shabad}

\address{I.E.Tamm Department of Theoretical Physics, P.N.Lebedev Physical Institute,
  Russian Academy of Sciences, Leninsky prospekt 53, Moscow 117924}

\ead{shabad@lpi.ru}

\begin{abstract}
  We consider the radial \Sch
with an attractive potential singular in
the origin. The additional continuum of states caused by the singularity,
that usually remain nontreatable, are shown to correspond
to particles, asymptotically free near the singularity (in the inner
 channel). Depending on
kinematics, they are either confined by the centre or may escape to
infinity (to the outer channel).

The orthonormality within
the continuum of confined states is established and the scattering phase
of the particle emitted by the centre and then reflected back to it is found.

For the deconfinement case a unitary 2$\times $2
S-matrix is found in terms of the Jost functions, and describes
transitions within and between the two channels. The volume elements
in the two channels
 are different.

   The two-channel situation is  analogous to the known behaviour of radiation
in the black hole metrics. We discuss the black hole essence of singularly
attracting centre for classical motion and  the relativity of time
inherent to this problem.

\end{abstract}


\submitto{\JPA}

\maketitle

\section{Introduction}

There are many reasons to be interested in studying singular potentials in quantum mechanics.
The most important ones lie in that these are inherent to the theory and cannot be merely
brushed aside referring to its physical nontreatability. For instance one cannot help
considering the case of supercritical nuclei with the charge $Z>137$ that creates effective
 singularity in the Dirac equation (see Berestetsky $et$ $al$ \cite{Ber}). When studying a
charged massive vector boson in a
magnetic field it is not consistent to confine oneself to considering only low fields below the
threshhold $B=m^2c^3/e\hbar$. Studying the Schr${\rm\ddot{o}}$dinger equation in the region of
 purely imaginary angular momentum is important in the framework of the Regge theory, especially
in proving
the Mandelstam  dispersion relations with respect to the scattering angle cosine (DeAlfaro and
Regge \cite{Reg}). The attractive singular centrifugal term that appears for imaginary angular
 momentum
 causes the fall down onto the centre  already at the free (potentialless) level.
The order-to-order growing divergencies that accompany the perturbation expansion in the Lippmann-
Schwinger equations in  singular attractive case
 are believed to model the nonrenormalizability in quantum field theory
\cite{Gut}, especially the growing divergencies in the ladder Bethe-Salpeter
equation. There are experimentally probed problems \cite{Den} of a
charged straight wire with a magnetic field applied along it or with the
Aharonov-Bohm potential, the both problems possessing \cite{Hag} the $-\beta/r^2$ singularity in
the Schr$\rm\ddot o$dinger equation (see \cite{Andr, Skar}).
We have listed these cases to illustrate the simple idea that consideration of singular potential
is inevitable.

The singular potential problem in quantum mechanics attracted essential efforts during,
 perhaps, half a century. In spite of a mathematical advancement achieved by applying the
 theory of  a self-ajoint extension of the Hamiltonian \cite{Akh, Pop}
(see Audretsch $et$ $al$ \cite{Andr, Skar} for the most recent results in this direction),
we cannot state that a
satisfactory physical description and understanding of the situation is developed of the fall
down onto the center phenomenon, peculiar to the singular potential problem. Nevertheless, at
least one lesson should be tought from these studies: there appears an extra continuum of states,
whose wave functions decrease at infinity. (These are sometimes inadequately called a continuum of
bound states).

In the present paper we study the radial \Sch and deal with the special case, mentioned above,
when the singularity
is caused by the centrifugal term $\lambda^2/r^2$ with the value $\lambda^2=l(l+1)+1/4$ (where
$l$ is the angular momentum) taken negative. The external potential is sufficiently well behaving
and real. We are going to fully exploit the appearance of extra degrees of freedom by giving a
direct physical interpretation to the continuum of states, presented by the asymptotic behaviour
of the  wave function near the origin $r=0$
\begin{eqnarray}
\psi(r)\asymp r^{\pm\rmi\rm{Im}\lambda+\case12}
=\exp{(\pm\rmi r_*\frac{\rm {Im}\lambda}{r_0})}r^\case12r_0^{\rmi{\rm Im}\lambda}
\end{eqnarray}
where $r_*=r_0\ln(r/r_0)$, and $r_0$ is a dimensional parameter. In our approach
this continuum is in the end responsible for the waves that behave like (asymptotically) free particles
 near the
origin, with Im$\lambda/r_0$ playing the role of their  momentum. These
waves are absorbed by the singular centre
or emitted by it . For negative energy $k^2<0$
they form a standing wave in the vicinity of the origin, no  overall probability
being
absorbed/emitted by the centre nor escaping to infinity. This is a regime, that
can be thought of as
confinement and is described as purely elastic process in the inner channel,
 when particles emitted from the centre are fully scattered back onto it.
When $k^2>0$ a two-channel
regime occurs, i.e. the particles, emitted by the center,
 not only are reflected back to the centre, but also partially escape to
infinity. Correspondingly, particles,
incoming from infinity, are partially reflected back into infinity and partially
penetrate into the inner region to become asymptotically free in the vicinity of
the centre (and be absorbed by it).

To the best of our knowledge the existing papers, including that of Alliluev
\cite{Alliluev} , where a set of waves, diverging from and converging to the
centre were considered, deal with the universal measure $\rmd r$ that serves
the scalar product both for $r\to\infty$ and $r\to 0$. The key difference of
our approach is that we show that another measure should be used, which is
singular  near the origin. Its use makes the norm of the oscillating in the
origin  wave function $linearly$ $divergent$. This fact is definitely necessary
to make it possible to interpret such wave functions as corresponding to
particles that are free in the origin and have at their disposal a sufficiently
ample volume
element in the centre to be such. This is analogous to the linear divergency
of the norm of the wave function of an ordinary free particle, infinitely remote
from the centre.

The above picture is similar to the behaviour
of scalar, electromagnetic, spinor and gravitational waves in the gravitational
field of a  spinning and nonspinning black hole, with particles disappearing
 behind the event horizon (see the book by Chandrasekhar \cite{Chandra} and the
  references to original works by himself, Zel'dovich, Starobinsky, Teukolsky
and others  therein).

The analogy of the singular centre and a black hole is, perhaps, more profound
than just the above two-channel similarity . In section \ref{Zeno} we discuss
the classical motion in an
 attractive $-\beta/r^2$ potential and argue that the character of the {\it unbounded}
 motion towards the singular centre proposes that the latter may be associated
 with a sort of preEinstein black hole, i.e. the one, not depending on the
 General Relativity Theory (GRT), but showing a relativity of time and two
different time scales, one of which is singular.

 In the rest of the paper we concentrate in the quantum mechanical case.

 In section \ref{Basic} the basic solutions of the radial
 \Sch and the corresponding Jost functions are listed.

In section \ref{Four} various two-channel
  and one-channel regimes are described in four domains (called {\it sectors})
  of the momentum and angular momentum values. This description is summarized
  in Table 1.

In subsections \ref{Bound}, \ref{Scattering} the ordinary bound states
 and elastic scattering sectors are briefly reviewed. In subsection \ref{Confined}
the confining sector is studied. The behaviour of the confined particle is
  described by a one-dimensional \Sch obtained by mapping the origin $r=0$
 to  the negative infinity $ r_*=-\infty$. The effective potential in the
    transformed \Sch disappears at $r_\ast =-\infty$ and exponentially
   grows for $r_*\to \infty$, thus providing the locking of the particle in the
   inner  space (half-bounded motion to and from the centre). The exponential
   growth of the locking potential is the consequence of the singularity
   in the initial \Sch.

The orthonormality condition for confined states is formulated referring to a scalar product
  with a new measure, singular in the origin: $\rmd r/{r^2}$. This measure appears
  naturally in the transformed \Sch and provides the necessary divergency of the norm
  of the wave function near $r=0$.

   In subsection \ref{Two}  the two-channel regime is studied. To describe it a $2\times2$
    $S-$matrix   ( $cf$ \cite {Chandra} for the black hole case) is obtained,
    whose unitarity $SS^\dagger=1$ follows from
   the properties of solutions of the \Sch, and reflects the overall probability
   conservation, as well as transformation laws of the Jost functions under
   the complex conjugation - the time reflection. The $S$-matrix contains three
   angles: two scattering phases and one extra angle, responsible for the mixing
   of the two channels - the inelasticity angle. The three angles are expressed
   in terms of three independent Jost functions.

    It is instructive, perhaps, for
   axiomatic quantum field theory, that the $S$-matrix unitarity cannot be
   formulated in terms of states, free at infinity, alone. The states,
   free near the origin, should be included.

   The orthogonality for the two-channel scattering states is established with a measure
that tends to $\rmd r$ for large distances from the centre and to $r_0^2\rmd r/r^2$ for small.
For properly arranged orthogonal enelastic scattering states the relation
    $k=-\rm {Im} \lambda/{r_o}$ holds (Im$k=$Re$\lambda=0$), which looks like a quantum number
conservation condition, if one understands $r _0$ as a conventional border between two channels.
  Finally, some prospects of future work are mentioned.

In concluding section \ref{Conclusion} the general insight into results is sketched.

\section{Fall down onto a centre as an unbounded motion in classical mechanics. PreEinstein black hole }
\label{Zeno}
In this section we confine ourselves to classical consideration of the central-symmetric motion in a potential
 that behaves in an attractive singular way near the origin $r=0$
 \begin{eqnarray}\label{pot}
                   U(r)=-\frac{\beta}{r^2}.
\end{eqnarray}
It dominates over the centrifugal barrier when
\begin{eqnarray}\label{beta}
                   \beta >\frac{L^2}{2m},
\end{eqnarray}
where $m$ is the mass and $L$ the angular momentum of a probe particle. When (2) is fulfilled,
the origin becomes a classically admissible point, since the square root
\begin{eqnarray}\label{R}
R=[(E-U(r))\frac{2}m-\frac{L^2}{m^2r^2}]^{1/2}
\end{eqnarray}
where $E>0$ is the energy, is real at $r=0$ ( the perihelium point corresponds to a complex
value of $r$). Then the motion is "unbounded" near the origin in the sense that the particle
turns around the centre infinite number of times before it reaches it following the logarithmic
spiral. This is seen from the expression for the angle differential
 \cite{Landaumech}

\begin{equation}\label{4a}
\rmd\phi=R^{-1}\frac{\rmd r}{r^2}\frac{L}m,
 \end{equation}
which leads to a logarithmic divergency in the origin $ \phi=-(\ln r)L
(2\beta m-L^2)^{-1/2}$. On the other hand, the time $t$, nesessary
 for the particle to reach the centre, and the distance $s$, which it passes before it
 does so, are both finite, since their differentials
 \begin{equation}\label{4b}
\rmd t=\rmd rR^{-1},\qquad \rmd s=r\rmd\phi
\end{equation}
are integrable in the point $r=0$. (Note that in the bordering case
 $\beta=L^2/{2m}$ the angle
 $\phi$ diverges like $L/r\sqrt{2Em}$, and $s$ also diverges
like -ln$r(L/\sqrt{2Em})$, while the time remains finite). If a potential
$U_1(r)$ behaves near the origin like (\ref{pot}) and is positive in some finite
region, there is an
 apohelium $r=r_{\rm max}$ to be found from the equation $U_1(r_{\rm {max}})=E$,
 and the trajectory is half-bounded: the particle may fall down onto the centre
 but cannot escape to infinity. On the other hand, there is also a half-bounded
 motion outside, provided that  $U_1(r)\to 0$ when $r\to\infty$. The situation when the
  motion is unbounded both at zero and at infinity is also possible ( for
  instance, if $U_1(r)<0$ everywhere). The same character of motion is peculiar
  to the quantum case, as it will be demonstrated in the next section. The main
  difference is that in quantum mechanics  the particle splits into reflected
  and transmitted parts.

  Now we are going to discuss what the spiral motion with infinite wending
  around the centre may have to do with the black hole. The black hole
  in GRT is characterized by the fact that there exist different
  ideas of $time$ held by different observers. The stationary (external)
  observer finds that the time necessary for a particle attracted by a
  black hole to disappear behind the horizon is infinite, whereas a comoving
  (internal) observer would spend only a finite (proper) time to reach the black hole.
  The possibility that different $times$ may exist is the manifestation of the principle of
  relativity of time, that reads that the {\it time} appears no sooner than
  means are pointed of how to make counts of it. The relativity of $time$
  is best pronounced within Einstein's theory of relativity, especially after
  the limitation on the maximum speed of a signal is imposed. It is certain,
  however, that the relativity of $time$, understood as stated above, is less
  special and exists beyond Einstein's relativity theory.

We advocate the theory that the famous Achilles-and-Tortoise paradox,
invented by Zeno of Elea as early as the fifth century BC, may be understood as
  introducing the relativity and singularity of {\it time}. Let us imagine, that
 an observer number 1
   does not have at his disposal another physical process for defining $time$
  except the Achilles-Tortoise pursuit. So he makes $time$ count each time
Achilles covers  the distance that separated him from Tortoise at the moment
of the previous count. Observer number 1 will make infinite number of counts
before Achilles  catches up with Tortoise. The clock of this observer will
indicate an infinite $time$. On the
  other hand, an observer number 2 who may use other physical processes, like planet
  (bounded) motion, pendulums, atomic transitions $etc$. to define segments of
  $time$ taken as equal, would state that Achilles catches up with Tortoise
  after only finite $time$ passes. The {\it time} of  observer number 1
 turns into infinity, {\it i.e.} it  posesses a singularity as a function of
the {\it time} of observer number 2 in a finite point of the latter. In this respect
 we say that Zeno's paradox introduces the situation of a sort of preEinstein black hole.

  Let us return to the singular centre ( \ref{pot}), ( \ref{beta}). Define
  observer number 1 as the one who does not have at his
   disposal any bounded motion to be used to produce  $time$ intervals, taken
as equal,   only the (unbounded) process of the fall down onto the centre.
 By definition, the only way for him to measure $time$ is by counting revolutions
   around the centre. These are infinite in number. Therefore, this observer
   would state that infinite time passes before the centre is reached, whereas the
 time obtained by integrating d$t$ in (\ref{4b}) is finite.

   The contents of the next sections do not depend directly on the consideration
   of the present section \label{Zeno}. What is remarkable is that in our classical
   consideration time and angle change their roles in the falling down onto the
   centre process: time is finite, whereas angle is infinite. On the contrary,
   for the motion, unbounded at infinity, the (scattering ) angle is finite
   and the time, necessary to reach infinity , is infinite. In quantum case we
   shall see in the next sections that the quantities, canonically conjugated
   to time and angle, i.e. the energy and the angular momentum (squared) also
   change their roles near the origin, leading to the change of the volume
   elements (the metrics).

  \section{Basic definitions}
\label{Basic}

In the rest of the paper we study the radial \Sch corresponding to spherically
symmetric case
\begin{eqnarray}\label{4}
H\psi(r)=k^2\psi(r)
\end{eqnarray}
\begin{eqnarray}\label{5}
H=-\frac{\rmd^2}{\rmd r^2}+\frac{\lambda^2-\case14}{r^2}+V(r),
\qquad 0\leq r<\infty,
\end{eqnarray}
where $k^2=E$ is the energy, and $\lambda$ is related to the
angular momentum quantum number as $\lambda=l+\case12$. The radial
part of the full 3D \Sch is connected with the solution of (\ref{4})
as $R(r)=\psi/r$. The potential $V(r)$ is real and good enough as not
to essentially affect the character of the solutions, as compared with
the free case of $V(r)\equiv 0$. Namely:

1) $V(r)$ decreases at $r\to\infty$ sufficiently fast so that the integral
\begin{eqnarray}\label{poten}
\int_c^\infty|V(r)|\rmd r<\infty
\end{eqnarray}
might converge at the upper limit

2) $V(r)$ is less singular in zero $r=0$ than the centrifugal term in
(\ref{4}), so that the integral
\begin{eqnarray}\label{potent}
\int_0^{c'}r|V(r)|\rmd r<\infty
\end{eqnarray}
converge at the lower limit, $c,c'>0$.

We shall give the parameters $\lambda$ and $k$ in (\ref{4}), (\ref{5})
real or purely imaginary values. Together with the reality of the potential
$V(r)$ this keeps coefficient functions in (\ref{5}) real, so that
if $\psi(r)$ is a solution to equation (\ref{4}) the complex conjugate
$\psi^*(r)$ is a solution of the same equation, too.

The singular character of the problem under
consideration is only due to the imagionarity of $\lambda$ and not to the
potential, that does not cause singularity according to (\ref{potent}).
For this reason in treating  our case many formal results may be borrowed
from the so-called Regge pole theory, dealing with analytical continuation
in the complex angular momentum plane, that was popular in the sixtieth-
seventieth, although no special emphasis was made at that time on the fact
that  there is a singular dynamics, underlying the formalism. The case of
$V(r)\equiv 0$ will be referred to as $free$, although this seemingly
 kinematic case is already far from being trivial and contains all the features
we are interested in.

In introducing basic definitions below we follow the book by DeAlfaro and
Regge \cite{Reg} . Consider two fundamental pairs of solutions of the \Sch.

\newcommand{\fmin}{f_{-\lambda,k}(r)}
\newcommand{\f}{f_{\lambda,k}(r)}
\newcommand{\fminn}{f_{-\lambda,-k}(r)}
\newcommand{\fk}{f_{\lambda,-k}(r)}
\newcommand{\pmin}{\phi_{-\lambda,k}(r)}
\newcommand{\pminn}{\phi_{-\lambda,-k}(r)}
\newcommand{\p}{\phi_{\lambda,k}(r)}
\newcommand{\pk}{\phi_{\lambda,-k}(r)}
The functions $\f$ and $\fk$ satisfy equation (\ref{4}) and have each a
single-term asymptotic behaviour at infinity
\begin{eqnarray}\label{8}
f_{\lambda,\pm k}(r)\asymp\exp (\mp\rmi kr),  \qquad r\to\infty.
\end{eqnarray}
The form (\ref{8}) is the leading term of a solution to equation (\ref{4}) in
the asymptotic region $r\to\infty$. The functions $\phi_{\pm\lambda,k}(r)$ satisfy
(\ref{4}) and have each a single-term asymptotic behaviour near the origin
\begin{eqnarray}\label{9}
\phi_{\pm\lambda,k}(r)\asymp r^{\pm\lambda+\case12},  \qquad r\to 0.
\end{eqnarray}
The solution $\f$ is even with respect to the change of sign of $\lambda$,
since neither the differential operator (\ref{5}), nor the boundary
condition (\ref{8}) depends on this sign: $\f=\fmin.$ Analogously,
$\p=\pk$.

There are linear relations between the solutions:
\numparts
\begin{eqnarray}\label{a}
\f=A\p+B\pmin,
\end{eqnarray}
\begin{eqnarray}\label{b}
\fk=C\p+D\pmin,
\end{eqnarray}
\begin{eqnarray}\label{c}
\p=E\f+G\fk,
\end{eqnarray}
\begin{eqnarray}\label{d}
\pmin=H\f+K\fk.
\end{eqnarray}
\endnumparts
Here the coefficients $A,B,C,D,E,G,H,K$ depend on the parameters
$\lambda$ and $k$, and are expressed in terms of the four Jost
functions $f(\pm \lambda,\pm k)$ , defined  as
\begin{eqnarray}\label{11}
f(\lambda,k)=W(\f,\p)=\f\frac{\rmd\p}{\rmd r}-\frac{\rmd\f}{\rmd r}\p,
\end{eqnarray}
where $W(f,\phi)$ is the Wronsky determinant of the two independent
solutions, as follows \cite{Reg}
\begin{eqnarray}\label{12}
\eqalign{A=-f(-\lambda,k)/{2\lambda},     \qquad &K=\rmi(\lambda/k)A,\\
E=\rmi f(\lambda,-k)/{2k},       \qquad &D=-\rmi(k/\lambda)E,\\
G=-\rmi f(\lambda,k)/{2k},       \qquad &B=\rmi(k/\lambda)G,\\
C=-f(-\lambda,-k)/{2\lambda},    \qquad &H=-\rmi(\lambda/k)C.}
\end{eqnarray}In obtaining (\ref{12}) the use was made of (\ref{11})
and the other two, trivial Wronskians
\begin{equation}
W(\p,\pmin)=-2\lambda, \qquad W(\f,\fk)=2\rmi k,
\end{equation}
as well as of the evenness of $\f$ with respect to $\lambda$, and of $\p$
with respect to $k$. From the Volterra integral equations for $\f,\p$,
equivalent to the differential equation (\ref{4}) with the boundary conditions
(\ref{8}), (\ref{9}), resp., one can get the complex conjugation properties
for the solutions \cite{Reg}
\begin{equation}\label{14}
\eqalign{(\f)^*=f_{\lambda^*,-k*}(k),\\
(\p)^*=\phi_{\lambda^*,k^*}(r).}
\end{equation}
The Jost function satisfies the relation \cite{Reg}
\begin{equation}\label{15}
f^*(\lambda,k)=f(\lambda^*,k^*\exp(-\rmi \pi)).
\end{equation}
In the free case , $V(r)\equiv0$, the above solutions and the Jost function are \cite{Reg}
\begin{equation}\label{16}
\eqalign{\phi^{(0)}_{\lambda,k}=\Gamma(\lambda+1)\left( \frac k{2}\right)^{-\lambda}
r^\case12 J_\lambda (kr),\\
f^{(0)}_{\lambda,k}=\left( \frac{\pi kr}{2}\right)^\case12
\exp(-\rmi \frac \pi{2}(\lambda+\case12)) H^{(2)}_\lambda(kr),\\
f^{(0)}(k,\lambda)=\left(\frac{\rmi k}{2}\right)^{-\lambda+\case12}(2/{\sqrt{\pi}})
 \Gamma(\lambda+1).}
\end{equation}
Here $J_\lambda (kr)$ and $H^{(2)}_\lambda(kr)$ are cylindric functions in the standard
notations, $\Gamma$ is the Euler gamma-function. The free solutions
 $f^{(0)}_{\lambda,k}(r), \phi^{(0)}_{\lambda,k}(r)$ posess the same asymptotic forms
(\ref{8}), (\ref{9}), resp., as $\f,\p$ owing to the properties (\ref{poten}),
 (\ref{potent}) of the potential.
\section{Four kinematic sectors}
\label{Four}

We are now in a position to describe four different sectors of the problem: sector I
(imaginary $k$, real $\lambda$), sector II (real $k$, real $\lambda$), sector III
(imaginary $\lambda$, imaginary $k$), and sector IV (imaginary $\lambda$, real $k$).
This consideration is summarized in Table 1.
\newcommand{\la}{\lambda}
\Table{\label{1}Four sectors I, II, III, IV of the singular Schr$\ddot{\rm o}$dinger
problem for various domains
of the momentum $k$ and the angular-momentum-related parameter $\la$}
\br
 &                &\it i m a g i n a r y \hspace{2.5mm} $k$   &\it r e a l \hspace{2.5mm} $k$\\
\mr
 &\it r\\
 &\it e            &Sector I                                   &Sector II\\
 &\it a\\
 &\it l            &Discrete manifold of\hspace{15mm}          &Continuum of\\
 &                 & BOUND STATES                              &ELASTIC SCATTERING states\\
 &$\la$                  &                                     &in the outer channel\\

 &\\
\mr
                           &\\
\it i &n\hspace{15mm}      &Sector III                         &Sector IV\\
\it m &a                   \\
\it a &r                     &Continuum of                     &Continuum of (inelastic)\\
\it g &y                      &CONFINED STATES:                &scattering states with\\
\it i- &                        &elastic scattering states      &TRANSITIONS BETWEEN \\
       &$\la$          &in the inner channel                 &TWO CHANNELS\\
                             &\\
\br
\endTable

Sectors I and II are quite ordinary, whereas sectors III and IV reveal the fall down
onto the centre features. For the sake of completeness and comparability
 we describe sectors I and II in the words that will be appropriate for
describing  sectors III and IV as well (see subsections \ref{Confined} and \ref{Two}).

\subsection{Bound states}
\label{Bound}
In sector I a discrete manifold of states may appear as usual.

For Re$\la\geq \case 12$, Im$\la=0$ we face repulsion in the vicinity of the origin,
the wave function $\p$ according to (\ref{9}) decreases fast enough to keep the radial
 part of the wave
function of the 3D \Sch $R(r)=\psi(r)/r$ finite (see Landau's and Lifshits' Quantum
Mechanics \cite{QM}).
Simultaneously, for Re$k=0$, Im$k>0$ the wave function $\fk$ is according to (\ref{8})
 decreasing and square-integrable near infinity $\parallel\fk\parallel_\infty<\infty$, where
\begin{equation}\label{18}
\parallel\psi\parallel_\infty=\int_{0}^\infty |\psi(r)|^2\rmd r,
\end{equation}
while the function $\f$ is not. Correspondingly, the physically acceptable states are given
 by zeros of the coefficient $E$ in (\ref{c}), $i.e.$ satisfy the equation $f(\la,-k)=0$
according to (\ref{12}). Its solutions, if any, make a discrete manifold of trajectories
Im$k_{n_r}({\rm Re}\la)$, labelled by an integer, $n_r=0,1...$ (For half-integral $\la_l=
l+\case 12,$ $ l=0,1...$ these are physical bound states, labelled by a couple of discrete
quantum numbers, $n_r$ and $l$.)
Finally, the wave function of the acceptable states is $\p=\fk$. As for the function $\pmin$, it cannot
provide the finiteness of $R(r)$ for any Im$\la=0$, Re$\la\geq\case12$, and is not to be used.

We are left with the interval $0\leq{\rm Re}\la<\case 12$, Im$\la=0$, not considered yet.
In this interval the interaction in (\ref{4}, \ref{5}) is already attractive, but not
strong enough to give rise to the fall down onto the centre \cite{QM}, since
  (\ref{9}) does not oscillate. The criterion of the finiteness of $R(r)$, used above,
is no longer applicable to select solutions: this function is infinite in $r=0$ for the
both functions $\phi_{\pm\la,k}(r)$. Generally the requirement that $R(r)$ should be finite
 in the origin lies beyond the context of the equation (\ref{4},\ref{5}) under study,
 is not natural from the formal point of view and should not be used at all. As for the norm
(\ref{18}), it is not indicative within the interval $0\leq{\rm Re}\la<\case 12$, since it
converges for the both functions  $\phi_{\pm\la,k}(r)$ there.

 As a matter of fact ( see the next subsections \ref{Confined}, \ref{Two}), a special measure,
which behaves like d$r/r^2$ near the origin, appears
naturally in the problem, to be used when calculating norms and scalar products.
Appealing to this measure we state that the norm
\begin{equation}\label{19}
\parallel\psi\parallel_0=\int_0^\infty |\psi(r)|^2\frac{\rmd r}{r^2}
\end{equation}
converges at the lower limit in sector I, for $\psi=\p$, Re$\la\geq 0$, while the same
integral (\ref{19}) for $\psi=\pmin$ diverges for Re$\la\geq 0$.

We conclude that physically accepted solutions $\p=\fk$ in sector I, that make - provided
they exist - a  dicrete set
specified by trajectories Im$k_{n_r}(\rm Re\la),$ are selected by the requirement that the
norm (\ref{18}) for them converges at the upper limit, and the norm (\ref{19}) at the lower.
(The convergences of (\ref{18}) at the lower, and of (\ref{19}) at the upper limit are then
automatically guaranteed). We shall add more concerning the orthonormality relations among
the acceptable solutions of sector I in subsection \ref{Orth}.
\subsection{Elastic scattering states}
\label{Scattering}
In sector II (real $k$, real $\la$) usual continuum of states that are free at infinity
 lies. Again, out of the two solutions $\p$, $\pmin$ according to (\ref{9}) only $\p$
has the norm (\ref{19}) converging at the lower limit for Re$\la\geq 0$. According to (\ref{c}), (\ref{8}) the
wave function $\p$ has two nondecreasing oscillating asymptotic forms at infinity,
an incoming and outgoing waves, of which none is better than the other. For the both of
them the norm (\ref{18}) $\parallel f_{\la,\pm k}(r)\parallel_\infty=\infty$ diverges
linearly with respect to the upper cutoff $R$, if the latter is introduced. The
normalization of scalar products by dividing over $R$ is used and ensures that the
particles stay mostly at an infinite distance from the centre.

The Wronsky determinant of two mutually complex conjugate solutions is (up to a factor of i)
the probability flux density to or from the centre.
\begin{equation}\label{20}
P(\psi)=\rmi\left(\psi\frac{\rmd\psi^*}{\rmd r}-\psi^*\frac{\rmd\psi}{\rmd r}\right).
\end{equation}
The normalization of the probability flux density remains at this stage nonfixed
and depends upon normalization of the appropriate wave function. Since there is
 no singularity between the points $r=0,\infty$, the probability flux density is
independent of $r$. In sector II the probability flux carried by the wave function
$\p$ disappears: $P(\p)=0$ already because this function is real near the origin
in accord with (\ref{9}). Thus, this function is responsible for elastic scattering
of the incoming wave by turning it into outcoming wave, with no probability being
transferred to the centre. The incoming and outgoing waves form a standing wave at
infinity with equal amplitudes and a phase difference, which is the scattering
phase.
The elastic scattering reduces to pure reflection with no transmission of the wave
towards the centre. The absence of penetration is due to the singular repulsive
(locking out) barrier (of kinematical nature) provided by the centrifugal term
$(\la^2-1/4)/r^2$ in (\ref{5}). The probability flux densities (\ref{20}) associated
 with the two functions $\f$ and $\fk$ forming $\p$ in view of (\ref{c}) are equal
in absolute value but differ in sign, the crossing terms being zero. The overall
probability, transferred to or from the infinity is also zero in virtue of the
constancy of the Wronsky determinant. "The infinity absorbs all it emits".

All these well known facts will find their
close analogy in the situation characteristic of sector III, to be described in
the next subsection \ref{Confined}.

\subsection{Continuum of confined states}\label{Confined}
In sector III of Table 1, wherein Re$\la=0$, Re$k=0$ consider the wave
 function $\fk$ in (\ref{b}). For Im$k>0$ owing to (\ref{8}) this function is
exponentially
decreasing, as $r\to\infty$, and real. It hasthe  norm (\ref{18})
convergent at the upper limit $\parallel\fk\parallel_\infty<\infty$, and carries to/from the infinity
a zero probability flux (\ref{20}) $P(\fk)=0$. As for its two asymptotic
forms near the origin (\ref{9}), they both oscillate, none of them being any
worse or better than the other. These states make
a continuum, since there is no reason to require that any of the
coefficients $C$ or $D$ in (\ref{b}) should turn to zero.
 We insist that the equality in rights of the two
asymptotically oscillating solutions (\ref{9}) in sector III leads to
analogous interpretation as the same circumstance in sector II, $ i.e.$
they are responsible for free particles, this time incoming from or outgoing to
the centre.
\subsubsection{Transformed \Sch.}
\label{Transformed}

To make this interpretation explicit let us perform the  transformation
 of the variable in the \Sch (\ref{4}), (\ref{5})
\begin{equation}\label{21}
r_*=r_0\ln\frac r{r_0},
\end{equation}
where $r_0$ is an arbitrary dimensional parameter, accompanied with the
transformation  of the wave function
\begin{equation}\label{22}
\left(\frac{r_0}{r}\right)^\frac1{2}\psi(r)=\tilde{\psi}(r_*)
\end{equation}
or
\begin{equation}\label{23}
\tilde{\psi}(r_*)=\exp (-\frac{r_*}{2r_0})\psi(r_0\exp \frac{r_*}{r_0})
\end{equation}
The transformation (\ref{21}) maps  the origin $r=0$ to the minus infinity
$r_*=-\infty$. The transformation (\ref{22}) is intended to annihilate
the first-order derivative term in the resulting differential equation.
 With (\ref{21}), (\ref{22}) the \Sch (\ref{4}), (\ref{5}) turns into
\begin{equation}\label{(24)}
H_{\rm cent}\tilde{\psi}(r_*)=\frac{(\rm Im\la)^2}{r_0^2}\tilde{\psi}(r_*),
\qquad -\infty<r_*<\infty,
\end{equation}
\begin{equation}\label{25}
H_{\rm cent}=-\frac{\rmd^2}{\rmd r_*^2}-\exp\left(\frac{2r_*}{r_0}\right)
\left(k^2-V(r_0\exp\frac{r_*}{r_0})\right).
\end{equation}
 In what follows we use the tilde sign to denote the
transform (\ref{22}), (\ref{23}) of any solution of equation (\ref{4}), (\ref{5}).

The probability flux density (\ref{20}) - like any
other Wronsky determinant - remains form-invariant under the transformation
(\ref{22}), (\ref{21})
\[\rmi\left(\psi(r)\frac{\rmd\psi^*(r)}{\rmd r}-
      \psi^*(r)\frac{\rmd\psi (r)}{\rmd r}\right)=
\rmi\left(\tilde{\psi}(r_*)\frac{\rmd\tilde{\psi}^*(r_*)}{\rmd r_*}-
      \tilde{\psi}^*(r_*)\frac{\rmd\tilde{\psi} (r_*)}{\rmd r_*}\right).
\]
This invariance would be violated if we have chosen different scales in the $r-$
and $r_*-$ coordinates by using the transformation $r_*=r_*^0\ln (r/r_0)$ with
$r_*^0\neq r_0$ in place of (\ref{21}).

Bearing in mind the property (\ref{potent}) of the potential one sees that in
the limit $r_*\to-\infty$ the potential-containing term in (\ref{25})
may be omitted (as well as the kinematic term $k^2$). This is the property of asymptotic freedom near the singular centre.
The asymptotic form of \eref{(24)} is the free equation, with Im$\la/r_0$ playing
the role of asymptotic momentum in the inner channel,
\begin{equation}\label{26}
-\frac{\rmd^2}{\rmd r_*^2}\tilde{\psi}_{-\infty}(r_*)=\frac{(\rm Im\la)^2}{r_0^2}\tilde{\psi}
_{-\infty}(r_*).
\end{equation}
Its solutions are
\begin{equation}\label{27}
\tilde{\psi}_{-\infty}(r_*)=c_\pm\exp \left(\pm \rmi \frac{\rm Im\la}{r_0}r_*\right),
\end{equation}
where $c_\pm$ are arbitrary constants, and Im$\la\equiv\sqrt{(\rm Im\la)^2}>0$.
 This is in agreement with (\ref{9}).

Consider
\begin{equation}\label{28}
\tilde{f}_{\la,-k}(r_*)=\exp (-\frac{r_*}{2r_0})f_{\la,-k}\left(r_0\exp \frac{r_*}{r_0}\right).
\end{equation}This function
is a combination of two waves: the one incoming from the negative infinity of $r_*$,
and the other outgoing to the negative infinity of $r_*$ (the waves, emitted and absorbed
by the center $r=0$). This is seen from (\ref{b}) when written as follows
\begin{eqnarray}\label{29}
\fl\tilde{f}_{\la,-k}(r_*)=\exp (-\frac{r_*}{2r_0})\left[C\phi_{\la,k}
\left(r_0\exp \frac{r_*}{r_0}\right)+D\phi_{-\la,k}\left(r_0\exp \frac{r_*}{r_0}\right)\right]\nonumber\\
\lo      =C\tilde{\phi}_{\la,k}(r_*)+D\tilde{\phi}_{-\la,k}(r_*)\nonumber\\
\lo\asymp C\exp \left(\rmi \frac{\rm Im\la}{r_0}r_*\right)r_0^{\rmi\rm Im\la+\frac1{2}}+
D\exp \left(-\rmi\frac{\rm Im\la}{r_0}r_*\right)r_0^{-\rmi\rm Im\la+\frac1{2}},\\
\lo r_*\to -\infty\nonumber
\end{eqnarray}
which agrees with(\ref{27}).

The norm
$\parallel\tilde{\psi}\parallel_{-\infty}=\int_{-\infty}^\infty|\tilde{\psi}(r_*)|^2\rmd r_*$,
 associated with \eref{(24)}), is just the norm (\ref{19})
 (and not (\ref{18})). Taken for the transform of $\fk$, it diverges linearly with the lower
 cutoff $L$
\[
  \lim_{L\to\infty}\int_{-L}^\infty|\tilde{f}_{\la,-k}(r_*)|^2\rmd r_*=L(|C|^2+|D|^2),
\]
 as it should for the continuum of asyptotically free particles ($cf$ sector II). This implies
that the particle is located mostly at negative infinity of the variable $r_*$. Such is the
meaning to be given to the phrase: fall down onto the centre.

At the positive infinity of $r$ also $r_*\to\infty$ and $V\left(r_0\exp (r_*/r_0)\right)$
may be as usual neglected as compared to $k^2$ in (\ref{25}) due to the property (\ref{poten}).
\Eref{(24)}, (\ref{25}) becomes
\begin{equation}\label{30}
-\frac{\rmd^2}{\rmd r^2_*}\tilde{\psi}_\infty(r_*)
-k^2\exp\left(\frac{2r_*}{r_0}\right)\tilde{\psi}_\infty(r_*)
 =\frac{(\rm Im\la)^2}{r_0^2}\tilde{\psi}_\infty(r_*)
\end{equation}
In agreement with (\ref{8}), (\ref{21}), (\ref{22}) the asymptotic behaviour of (\ref{29})
for $r_*\to\infty$ is
\begin{equation}\label{31}
\tilde{f}_{\la,-k}(r_*)\asymp\exp \left(-r_0{\rm Im} k\exp \frac{r_*}{r_0}\right)\exp (-\frac{r_*}{2r_0}).
\end{equation}
This is the leading asymptotic term in the solution of (\ref{30}).
Remind that in sector III $k^2<0$, and we are considering Im$k>0$. Hence the second term in
the l.-h. side of (\ref{30})
is an infinitely growing positive (repulsive) potential, locking the particle away from the
outer world. It is just $k^2r^2$ and originates from the
$1/r^2$ singularity in the initial equation (\ref{4},\ref{5}).
It prevents the particle from escaping to the positive infinity $r\to\infty$
in the same way, as the infinite repulsive centrifugal term in sector II prevented it from
approaching the origin ($cf$ subsection \ref{Scattering}). Again, the locking potential in sector III is a
purely kinematic (kinetic energy) term that originates from the r.-h. side of (\ref{4}),
 whereas in sector II it was the centrifugal potential, forming the r.-h. side of \eref{(24)}.
The function (\ref{31}) for Im$k>0$ decreases, as it should.

Thus, $\tilde{f}_{\la,-k}(r_*)$
represents the process of elastic scattering of particles incoming from the negative infinity
 of $r_*$ ($i.e.$  from the origin
of $r$) with all the probability reflected inwards (towards $r=0$), nothing escaping to
positive infinity. In other words, one has purely elastic scattering process
in sector III, similar to the one in sector II, with the only difference that
 the asymptotically free particles  income and outgo from/to the origin $r=0$,
$r_*\to-\infty$.

In the asymptotic domain $r_*\to-\infty$ the solution (\ref{29}) may be presented as
\begin{equation}\label{32}
\tilde{f}_{\la,-k}(r_*)\asymp\left(S_{\rm{cent}}\exp \left[\rmi \frac{\rm Im\la}{r_0}r_*\right)
+\exp \left(-\rmi \frac{\rm Im\la}{r_0}r_*\right)\right]Dr_0^{-\rmi\rm Im\la+\frac1{2}},
\end{equation}
where the scattering  1$\times 1$ matrix (just a complex number) in sector III for the elastic
scattering from the centre back to the centre  is
\begin{equation}\label{33}
S_{\rm{cent}}=\frac C{D}r_0^{2\rmi\rm Im\la}=-\frac{f(-\la,-k)}{f(\la,-k)}r_0^{2\rmi\rm Im\la}.
\end{equation}
We have used (\ref{12}) in deriving the last equality in (\ref{33}).

Since in sector III the equaity $S_{\rm{cent}}S_{\rm{cent}}^*=1$ holds true due to (\ref{12}),
 (\ref{15}), the scattering matrix
(\ref{33}) can be presented in the form, analogous to the well-known elastic scattering matrix
in sector II
\[S_{\rm{cent}}=\exp (\rmi\delta).\]
In the free case, $V\equiv 0$, using (\ref{16}) one obtains for the scattering phase
$\delta=\delta_0$, where
\begin{eqnarray}\label{34}
\fl\delta_0(\la,k)=\pi+2{\rm Im}\la\ln\left(\frac{r_0{\rm Im}k}{2}\right)
+\arg \frac{\Gamma (1-\rmi\rm Im\la)}{\Gamma(1+\rmi\rm Im\la)}\nonumber\\
\lo=\pi+2{\rm Im}\la\left[\ln\left(\frac{r_0{\rm Im} k}{2}\right)+C_{\rm E}\right]\nonumber\\
\lo+2 \arctan ({\rm Im}\la)
+2\sum_{s=1}^\infty\left[\arctan\left(\frac{{\rm Im}\la}{s+1}\right)-\frac{{\rm Im}\la}{s}\right],
\end{eqnarray}
where $C_{\rm E}=0.577...$ is the Euler constant, and the series in the r.-h.side converges.
\subsubsection{Orthonormality of confined states.}
\label{OCS}

Now we follow the common procedure, applicable irrespective of whether sector III or II is
concerned, to demonstrate that the confined wave functions $\tilde{\eta}_{\la,-k}(r_*)$,
normalized as
\begin{equation}\label{35}
\tilde{\eta}_{\la,-k}(r_*)=\frac{{\rm Im}\la \sqrt{2}}{\sqrt{r_0}|f(\la,-k)|}
\tilde{f}_{\la,-k}(r_*),
\end{equation}
satisfy the  orthnormality condition, taken for coinciding $k$'s and different
or equal $\la$'s. The orthogonality for different $\la$'s is independently
and straightforwardly demonstrated in subsection \ref{Orth}.

Consider the function
\begin{equation}\label{36}
\Delta_L(\la,\la',k)\equiv\int_{-L}^\infty\tilde{\eta}_{\la,-k}(r_*)
\tilde{\eta}_{\la',-k}^*(r_*)\rmd r_*
\end{equation}
and its integral with a test function $T({\rm Im}\la'/r_0)$\
\begin{equation}\label{37}
\tau\left(\frac{{\rm Im}\la}{r_0}\right)=\int_0^\infty\Delta_\infty(\la,\la',k)T\left(\frac{{\rm Im}\la'}
{r_0}\right)\rmd\frac{\rm Im \la'}{r_0}
\end{equation}
as a limit $L\to\infty$, where $-L$ is the size of the box $(-L,\infty),$
 into which the system is placed. The solution (\ref{35}) in the box is
subject to two homogeneous boundary conditions. The first one
\newcommand{\bee}{\begin{eqnarray}}
\newcommand{\eend}{\end{eqnarray}}
\bee\label{38}
\tilde{\eta}_{\la,-k}(\infty)=0
\eend
is fulfilled due to (\ref{8}) or (\ref{31}). The second boundary condition
imposed at the other end of the interval
\bee\label{39}
\tilde{\eta}_{\la,-k}(-L)=0
\eend
is satisfied for dicrete values of Im$\la$. For large $L$ these
values are found from the equation that follows from (\ref{32}),
(\ref{35})
\bee\label{40}
-\frac{{\rm Im}\la_n}{r_0}L+\frac1{2}\delta(\la_n,k)=-n\pi+\frac\pi{2}
\eend
to be
\bee\label{41}
\frac{{\rm Im}\la_n}{r_0}=\frac{n\pi}{L}
\eend
with $n=0,1,2...\infty$ unless $\delta$ grows too fast when $\la\to 0$.
With (\ref{41}) in mind the integral (\ref{37}) is presented as the
limit at $L\to\infty$ of the Darboux sum (we take d$($Im$\la'/r_0)=\pi/L)$
\bee\label{42}
\tau\left(\frac{\pi n}{r_0}\right)=\lim_{L\to\infty}\sum_{n'=0}^\infty \Delta_L(\la_n,\la_{n'},k)
\frac\pi{L}T\left(\frac{\pi n'}{L}\right)
\eend
In calculating the limit $\Delta_L/L$ we take into account that only
the diverging part of $\Delta_L$ may give a finite contribution into it,
 and hence the integrand in (\ref{36}) should be replaced by its
 asymptotic form at $r_*\to-\infty$ according to (\ref{29}) (integral
(\ref{36}) converges at the upper limit $r_*\to\infty$ in sector III
due to (\ref{8})). With the use of (\ref{b}), (\ref{29}) one obtains
\bee\label{43}
\fl\lim_{L\to\infty}\frac{\Delta_L}{L}=\lim_{L\to\infty}\frac1{L}
\int_{-L}^\infty\tilde{\eta}_{\la,-k}(r_*)
    \tilde{\eta}_{\la',-k}^*(r_*)\rmd r_*
=\frac1{2|f(\la,-k)f(\la',-k)|}\nonumber\\
\times\lim_{L\to\infty}\frac1{L}\int_{-L}^c
\left[f(-\la,-k)f^*(-\la',-k)r_0^{\rmi{\rm Im}(\la-\la')}
\exp \left(\frac{\rmi{\rm Im}(\la-\la')}{r_0}r_*\right)\nonumber \right.\\ \left.
\lo+f(\la,-k)f^*(\la',-k)r_0^{\rmi{\rm Im}(\la'-\la)}
\exp \left(\frac{\rmi{\rm Im}(\la'-\la)}{r_0}r_*\right)\right]\rmd r_*,
\eend
where $c$ is any finite number. The difference between the integrals
in the first and second lines in (\ref{43}) is finite and gives vanishing
contribution into the limiting value of $\Delta_L/L$. Also for this reason we
 omitted the crossing terms $\exp (\pm\rmi{\rm Im}(\la+\la')r_*/r_0)$, since
$\la$ and $\la'$ are of the same sign and the sum $\la+\la'$ is nonzero.
As long as $\la\neq\la'$ the integral in (\ref{43}) is finite and the limit
(\ref{43}) disappears as $L\to\infty$. For $\la=\la'$  expression (\ref{43})
is unity in virtue of (\ref{15}). We conclude, that
\bee\label{45}
\lim_{L\to\infty}\frac{\Delta_L}{L}=\delta_{n,n'}.
\eend
Hence, (\ref{42}) becomes $\tau({\rm Im}\la/r_0)=\pi T({\rm Im}\la/r_0)$, which implies that
$\Delta_\infty$ in (\ref{37}) is $\pi$ times the Dirac $\delta$-function. Referring to
(\ref{36}) we finally get the orthnormality relation in sector III (Re$\la$, Re$\la'=0$,
Im$\la$, Im$\la'>0$, Im$k>0$, Re$k=0$)
\bee\label{46}
\Delta_\infty(\la,\la',k)=\int_{-\infty}^\infty\tilde{\eta}_{\la,-k}(r_*)
\tilde{\eta}_{\la',-k}^*(r_*)\rmd r_*=\pi\delta\left(\frac{{\rm Im}\la}{r_0}
-\frac{{\rm Im}\la'}{r_0}\right)
\eend

Note, that as long as the parameter Im$\la$ can be  viewed upon as a strength of
the singular attraction, the orthonormality relation (\ref{46}) expresses
spectral properties with respect to a "coupling constant". This may be interesting when
singular potentials other than $1/r^2$ are concerned.

To conclude this subsection let us transform the integral (\ref{46}) to the primary
variable. Using (\ref{21}), (\ref{22}), (\ref{35}) we obtain from
(\ref{46})
\bee\label{48}
\int_0^\infty\rho_{\la,-k}(r)\rho_{\la',-k}^*(r)\frac{\rmd r}{r^2}
=\pi\delta({\rm Im}\la-{\rm Im}\la'),
\eend
where
\bee\label{49}
 \rho_{\la,-k}(r)=\frac{{\rm Im\la}\sqrt{2}}{|f(\la,-k)|}\fk
\eend
Relation (\ref{48}) does not contain $r_0$. Note the singular measure
$r^{-2}$d$r$ that appears in (\ref{48}).

For the free case, $V(r)\equiv 0$, it follows from the first line of (\ref{34})
that (\ref{41}) is indeed the solution to equation (\ref{40}). After substituting
(\ref{16}) into (\ref{49}) we find the normalized wave function of a confined state
\bee\label{48a}
\rho^{(0)}_{\la,-k}(r)=r^\case12 K_{\rmi{\rm Im}\la}(r{\rm Im}k)
[{\rm Im}\la\sinh(\pi{\rm Im}\la)]^\case12\left(\frac 2{\pi}\right)^\case12
\eend
and (\ref{48}) takes the form
of a relation for McDonald functions with imaginary indices $K_{\rmi{\rm Im}\la}(r)$:
\bee\label{48b}
\frac 2{\pi}{{\rm Im}\la\sinh (\pi{\rm Im}\la})\int_0^\infty K_{\rmi{\rm Im}\la}(r)
K_{\rmi{\rm Im}\la'}(r)\frac{\rmd r}{r}
=\pi\delta ({\rm Im}\la-{\rm Im}\la')
\eend
Here, the orthogonality for Im$\la\neq{\rm Im}\la'$ may be directly seen from the formula 6.576.4
of Gradshteyn and Ryzhik \cite{Ryz} after one makes the power $\la$ tend to 1 in it and puts
$a=b$. The $\delta$-function is given rise to by the divergency of (\ref{48b}) near $r=0$. The
normalization may be checked using the asymptotic behaviour of McDonald function near $r=0$
following the example \cite{QM} of one-dimensional \Sch with the barrier reflection.
\subsection{Two-channel sector}
\label{Two}
In sector II the (elastic) scattering states were described by one function $\p$ out of the set
of two fundamental solutions $\phi_{\pm\la,k}$, while the other, $\pmin$ was physically useless.
In sector III, again, only one function, $\fk$ was responsible for elastic scattering, while its
partner from the fundamental couple, $\f$ was meaningless. Correspondingly, the scattering
matrices in these sectors were unitary  $1{\times}1$ matrices, $\it {i.e.}$ just complex numbers
with their absolute values equal to unity. One had a single scattering phase in each sector.

In sector IV, where $\la$ is imaginary and $k$ real, both solutions from each fundamental
couple  are meaningful and represent waves incoming from and outgoing to infinity, as
(\ref{8}), and incoming from and outgoing to the origin, as (\ref{9}) ( or from/to the
negative infinity of $r_*$ (\ref{21}), as (\ref{27})). Any linear combination (13) of two
meaningful fundamental solutions is again a meaningful solution. Two $2{\times}2$ mutually
inverse matrices transform two fundamental sets of two solutions according to (13).
\subsubsection{$2{\times}2$ scattering matrix.}
\label{2x2}

To calculate the probability flux density $P(\p)$ brought from the centre $r=0$ in the physical
 state described by the wave function $\p$ it is sufficient to substitute (\ref{9})
into (\ref{20}). This gives $P(\p)={\rm Im}\la$. The same quantity (mind the probability
conservation, $i.$$e.$ the Wronskian (\ref{20}) independence of $r$) is to be obtained,
 provided we use
the r.-h.side of (\ref{c}) in the asymptotic form near $r\to\infty$
\bee\label{50}
\p=E\exp (-\rmi kr)+G\exp (\rmi kr)
\eend
and substitute this into (\ref{20}). The crossing term  $\sim EG$ vanishes, while the first term
gives the contribution $-k|E|^2$ responsible for the influx of probability from infinity, and the
second term contributes as $k|G|^2$, which is the probability carried away to infinity. Therefore
\bee\label{52}
 \frac{|G|^2}{|E|^2}-\frac{{\rm Im}\la}{k|E|^2}=1.
\eend
For $\rm Im\la<0, k>0$ the flux density $P(\p)$ is negative and the wave $\p$ is absorbed by
the centre. Then relation (\ref{52}) states that the sum of reflection and transmission
coefficients is unity. For Im$\la>0, k>0$ the centre emits, and the reflection coefficient
$|G|^2/|E|^2$ in (\ref{52}) is greater than unity ($cf$ the superradiation of the Kerr
black hole \cite{Chandra}).

Analogously, in the state respresented by the wave function $\fk,$ $k>0$ the probability flux density
(\ref{20}) emitted outwards is $P(\fk)=k$. By equating this to the same quantity calculated using
 (\ref{b}) and (\ref{9}) we obtain the probability conservation relation
\bee\label{53}
k={\rm Im}\la |C|^2-{\rm Im}\la|D|^2,
\eend
where the first term in the r.-h.side describes the probability flux emitted by the centre,
when Im$\la>0$, or absorbed by it, when Im$\la<0$, while the second term $vice$ $versa$.
Thus, for Im$\la$ negative, after normalizing to the wave incoming from the centre, we obtain
from (\ref{53})
\bee\label{54}
1=\frac{|C|^2}{|D|^2}-\frac k{{\rm Im}\la|D|^2}.
\eend
This reads: the coefficient of reflection of the wave emitted from the origin back into the origin
plus the coefficient of transmission of the emitted wave to infinity is unity. For positive Im$\la$
the relation that expresses the same physical statement is obtained from (\ref{53}) by dividing
it over Im$\la |C|^2$, which is again the normalization to the emitted wave. This relation is
\bee\label{55}
1=\frac{|D|^2}{|C|^2}+\frac k{{\rm Im}\la|C|^2}.
\eend
If we assume $k<0$ in (\ref{54}), (\ref{55}) we obtain that the reflection coefficient in the inner
channel is greater than unity, which means the internal superradiation due to the fact that
there is the overall incoming probability flux from the infinity.

In what follows we consider only negative Im$\la$ and positive $k$ for definiteness.

Represent (\ref{c}), (\ref{b}) in the form
\bee\label{56}
\f=-\frac G{E}\fk+\frac 1{\sigma E}\sigma\p,
\eend
\bee\label{57}
\sigma\pmin=\frac {\sigma}{D}\fk-\frac C{D}\sigma\p,
\eend
where the parameter $\sigma$ is introduced to appropriately normalize the wave
function $\p$
\[
\sigma=\left(\frac k{{\rm Im}\la}\right)^\case12.
\]
Define a ${2\times}2$ scattering matrix $S_{ij}, i,j=1,2$. Let $S_{11}=-G/E$ be
the coefficient that relates the wave, reflected to infinity, contained in the
asymptotic behaviour of $\fk$ in (\ref{56}), to the wave incoming from infinity,
contained in $\f$. Let, further, $S_{12}=(1/\sigma E)$ be the
coefficient that relates the wave transmitted to the origin, contained in $\sigma\p$
in (\ref{56}), to the wave
incoming from infinity, contained in $\f$. Also $S_{22}=-C/D$ relates the wave
reflected to the origin, contained in $\sigma\p$ in (\ref{57}), to the wave incoming
from the origin, contained in $\sigma\pmin$, while $S_{21}=(\sigma /D)$
relates the wave transmitted to infinity, contained in $\fk$ in (\ref{57}), to
the wave incoming from the origin $\sigma\pmin$. Using (\ref{12}) we can write (\ref{56}),
(\ref{57}) with the help of the $S$-matrix defined above:
\bee\label{58}
\left(\begin{array}{c}
\f
\vspace{7mm}\\
\sigma\pmin\end{array}\right)=
\left(\begin{array}{cc}
\frac {f(\la,k)}{f(\la,-k)}                &-\frac {2\rmi \sqrt{k{\rm Im}\la}}{f(\la,-k)}
\vspace{7mm}\\
\frac {2\rmi \sqrt{k{\rm Im}\la}}{f(\la,-k)}   &\frac {f(-\la,-k)}{f(\la,-k)}
\end{array}\right)
\left(\begin{array}{c}
\fk\vspace{7mm}\\\sigma\p\end{array}\right)
\eend
The nondiagonal part of the $S$-matrix in (\ref{58}) is antisymmetric: $S_{ij}=-S_{ji}$
for $i\neq j$. The unitarity of the $S$-matrix is
\bee\label{59}
SS^\dagger=1,
\eend
where $S^\dagger$ is Hermitean conjugate to $S$. The two diagonal equations in (\ref{59}) are just the relations  (\ref{52}) and  (\ref{54}),
that follow from the probability conservation. The two off-diagonal equations in  (\ref{59})
follow from the complex conjugation rules  (\ref{15}) for the Jost functions.

Note, that the $S$-matrix in  (\ref{58}) contains only three out of the four Jost functions
$f(\pm \la,\pm k)$. On the other hand, the Hermitean conjugate matrix $S^\dagger$ contains via
 (\ref{15}) another triple of the Jost functions, so that the unitarity property  (\ref{59})
implies a relation between the four Jost functions in sector IV. Such a relation is known
\cite{Reg}
\bee\label{60}\left|\begin {array}{cc}
     f(\la,-k)    & f(-\la,-k)\\

      f(\la,k)    &f(-\la,k)
\end{array}
\right|
=-4k\rm Im \la
\eend
and equivalent to (\ref{52}), (\ref{54}).

The most general form of a unitary matrix with antisymmetric nondiagonal part is
\bee\label{61}
S=\rme^{\rmi\delta_2}
\left(\begin{array}{cc}
\exp (\rmi\delta_1)\cos\alpha    &\sin\alpha\vspace{7mm}\\
-\sin\alpha                     &\exp (-\rmi\delta_1)\cos\alpha
\end{array}\right)
\eend
By comparing this with (\ref{58}) we find
\bee\label{62}
\delta_1=\arg f(\la,k), \hspace{10mm}
\delta_2=\arg f(\la,-k)+\frac\pi{2}, \nonumber\\
\tan\alpha=\frac{2\sqrt{-k{\rm Im}\la}}{|f(\la,k)|}.
\eend
The two scattering phases $\delta_1,\delta_2$ and the channel-mixing nonelasticity angle
$\alpha$ are real in sector IV, Im$\la<0,k>0$. The element $S_{11}$
in (\ref{58}) is
the usual form of the $S$-matrix in sector II, where the scattering of the particles
incoming from infinity is elastic: $|S_{11}|=1$ due to the relation$f(\la,k)=f^*(\la,-k)$,
valid for real $\la$, real $k$ (see (\ref{15})). The element $S_{22}$ in (\ref{58}) should
be compared with the $S_{\rm cent}$-matrix in sector III (\ref{33}), where the scattering
of particles emitted by the centre back to the centre is elastic: $|S_{22}|=1$ owing to
the relation $f(\la,-k)=f^*(-\la,-k)$, valid for imaginary $\la$, imaginary $k$,
(see (\ref{15})).

The above definition of the $S-$matrix is subject to an arbitrariness. Without affecting
the unitarity and the meaning of the $S-$matrix elements we can change the normalization
in (\ref{56}), (\ref{57}) by multiplying $\sigma$ by a unit length complex number, say
$\exp (\rmi \delta_3)$. Correspondingly, the nondiagonal elements $S_{12}$ and $S_{21}$ in
(\ref{61}), (\ref{58}) are multiplied by $\exp (-\rmi\delta_3)$ and $\exp (\rmi\delta_3)$,
resp. with the phase $\delta_3$ arbitrary. This tells us that the antisymmetricity of the
$S-$matrix we used for a particular choice of $\delta_3$ is not an invariant property.

The partial transmission of the wave incoming from infinity to the centre is considered as
a transition from the outer channel, formed by the environment of the infinitely remote
point, to the inner channel, formed by the environment of the origin. $Vice$ $versa$, the
 partial escaping of the wave, emitted by the centre to infinity is considered as the
reciprocal
transition between the two channels. Each interchannel transition means absorption in the
sense that the probability partially leaves the initial channel and is lost for it. For
this reason in sector IV the $S$-matrix elements $S_{11}$ and $S_{22}$ are no longer unit
length complex numbers, and the unitarity can only be formulated with the inclusion of the
elements $S_{12}, S_{21}$ responsible for transitions between the two channels as it was
 done above.

It would not be appropriate to try to take into account the nonelasticity
of the scattering process in sector IV by analytic continuation with respect to $k$ or
$\la$ from any of the elastic sectors II or III. The analytic continuation makes the
corresponding scattering phase complex but is unable to create the lacking phase
and the nonelasticity angle: a description of the system with a greater number of
degrees of freedom cannot be achieved by mere analytic continuation.

For the free case, $V(r)\equiv 0$, one readily gets from (\ref{62}), (\ref{16})
\bee\label{61a}
\delta_1^{(0)}=\frac\pi{4}-{\rm Im}\la\ln\frac k{2}+\frac 1{2}\arg\frac{\Gamma(1+\rmi{\rm Im}\la)}
{\Gamma (1-\rmi{\rm Im}\la)}, \hspace{10mm}\delta_2^{(0)}=\delta_1^{(0)}-\frac\pi{2},\\
\tan\alpha^{(0)}=\left(\rme^{-2\pi{\rm Im}\la}-1\right)^\frac 1{2}.
\eend
The transmission coefficient in (\ref{52}) is
\bee\label{61c}
T^{(0)}=\frac{-{\rm Im} \la}{k|E^{(0)}|^2}=\frac{-4{\rm Im}\la}{|f^{(0)}(\la,-k)|^2}=1-\rme^{2\pi{\rm Im}\la}.
\eend
The transmission vanishes, $\alpha^{(0)}=0$,  $T^{(0)}=0$  for  Im$\la=0$, and is a
maximum, $\alpha^{(0)}=\pi/2$,  $T^{(0)}=1$  for Im$\la =-\infty$. Expression (\ref{61c})
is in agreement with the
absorption cross section of \cite{Alliluev}.
\subsubsection{Orthogonality and measure.}
\label{Orth}
By combining any two solutions $\psi_{1,2}$ of
different Schr$\ddot{\rm o}$dinger equations (\ref{4}), corresponding to different values
of $k^2$ and $\la^2$, and integrating by parts we obtain
\bee\label{63}
\int_0^\infty\left[\frac{\la_1^2-\la_2^2}{r^2}-(k_1^2-k_2^2)\right]\psi^*_1(r)\psi_2(r)\rmd r
\nonumber\\=
\left.\left(\psi^*_1(r)\frac{\rmd\psi_2}{\rmd r}-\psi_2(r)\frac{\rmd\psi^*_1(r)}{\rmd r}\right)
\right|_0^\infty.
\eend
This holds true both with and without the complex conjugation sign over $\psi_1(r)$. To avoid
a possible misunderstanding, stress that the Wronskian-like form in the r.-h.side of (\ref{63})
is not $r$-independent, since  $\psi_{1,2}(r)$ are solutions of differential equations with
different coefficients in them.

We were/are interested in the functions $\psi_{1,2}(r)$ that:
\begin{itemize}
\item decrease at the both ends of the interval $0\leq r<\infty$. These make the discrete set of
functions $\psi_{1}(r)=\phi_{\la_{1},k_{1}}(r)=f_{\la_{1},-k_{1}}(r)$ and
$\psi_{2}(r)=\phi_{\la_{2},k_{2}}(r)=f_{\la_{2},-k_{2}}(r)$ of sector I
 (see subsection \ref{Bound})
\item decrease at the upper end , $r=\infty$, and turn to zero at the lower end
when the latter is understood as the limit of the left box end coordinate $r_*=-L$,
when $L\to\infty$ $ (r=r_0\exp(-L/r_0)\to 0)$. These are the functions
$\psi_{1,2}(r)=f_{\la_{1,2},-k}(r)$, $k_1=k_2=k$, corresponding to the continuum of
confined states of sector III (see subsection \ref{Confined})
\item turn to zero at the lower end, $r=0$, and at the upper end, provided that the latter
 is understood as the limit of the right box wall coordinate $r=R$, when $R\to\infty$. These are
 the wave functions $\psi_{1,2}(r)=\phi_{\la,k_{1,2}}(r)$, $\la_1=\la_2=\la$ of the
continuum of states of sector II (see subsection \ref{Scattering})
\item turn to zero at the both ends of the interval when these are both understood as limits
of the left and right box wall coordinates $L,R$ tending to infinity
($r_*=-L\to-\infty$, $r=R\to\infty$). These are the functions
 $\psi_{1,2}(r)$ formed by linear combinations of the functions
$\phi_{\la_{1},k_{1}}(r), \phi_{\la_{2},k_{2}}(r)$
and $f_{\la_{1},k_{1}}(r), f_{\la_{2},k_{2}}(r)$ that belong to the continuum of the
two-channel scattering states of sector IV, we are discussing in the present subsection.
\end{itemize}

In all these cases the r.-h.side of (\ref{63}) vanishes, provided that the integration limits
in the l.-h.side are understood as stated above. Now, the orthogonality relations that
 follow from (\ref{63}) are different in different cases and sectors. In sector I for bound
states and in sector II for elastic scattering states we are interested in orthogonality of states
with different energies $k_1^2\neq k^2_2$ and coinciding angular momenta. So we put $\la_1^2=
\la_2^2$ in  (\ref{63}) and see  that such states are orthogonal with the measure $\rmd r$
taken in the scalar product. The states with equal momenta squared $k_1^2=k_2^2$ but different
angular momenta $\la_1^2\neq\la_2^2$ belonging to the discrete specrtum of sector I or to the
continuum of confined states of sector III are orthogonal with the measure $\rmd r/r^2$ like in
(\ref{48}).

In sector IV, let us rearrange the set of solutions of the \Sch labelled by two parameters
$k$ and Im$\la$ into another two-parametric set labelled by $r_0$ and $k$, so that
\bee\label{64}
{\rm Im}\la=-r_0k.
\eend
Then  (\ref{63}) takes the form ( we assume the zero boundary conditions
imposed as discussed above)
\bee\label{65}
(k_1^2-k_2^2)\int_0^\infty\psi^*_1(r)\psi_2(r)\left(\frac{r_0^2}{r^2}+1\right)\rmd r=0.
\eend
 Here $\psi_1(r)$ refers to the set of quantum numbers  $ k=k_1, \la=-\rmi r_0 k_1$, whereas
 $\psi_2(r)$ to the set $ k=k_2, \la=-\rmi r_0 k_2$. Hence (\ref{65}) implies the orthogonality
 of solutions with different $k^2$'s (simultaneously, with different $\la^2$'s) with the measure
\bee\label{66}
\left(\frac{r_0^2}{r^2}+1\right)\rmd r
\eend
that tends to $\rmd r$ for $r\to\infty$ to provide the possibility of free life of particles
 at infinity, like in sector II, and tends to $(r_0^2/r^2)\rmd r$ for $r\to 0$ to provide the
freedom in the vicinity of the centre, like in sector III.

Relation (\ref{64}) should be substituted into (\ref{58}), (\ref{62}) and other equations of
subsection \ref{Two}.

From (\ref{66}) the meaning of the dimensional parameter $r_0$ becomes clear. It characterizes
the separation between the outer and inner worlds in the sense that: the greater $r_0$, the
greater the weight of the first term in the measure (\ref{66}), in other words, the farther
 the inner world spreads. So $r_0$ may be thought of as the size of the singular centre.
The parameter $r_0$ represents the barrier that reflects particles inwards and outwards.
Relation (\ref{64}) is a sort of quantum number conservation law at the "border": the angular
momentum, labelling the states in the inner world, is equal to the momentum, labelling the
states in the outer world, multiplied by the radius of the border.

Remarkably, with the parametrization (\ref{64}) the eigenvalues in the r.-h.sides of equations
(\ref{4}) and (26) coincide.

The final remark is in order. Let us represent the radial \Sch (\ref{4}, \ref{5}) in the form
 of a special eigenvalue problem
\bee\label{67}
\left(-\frac{\rmd^2}{\rmd r^2}+V(r)\right)\psi (r)=\kappa (r)\psi (r)
\eend
where $\kappa (r)$ is not a constant eigevalue, but a function:
\bee\label{68}
\kappa (r)=k^2-\frac{\la^2-\case14}{r^2}.
\eend
According to the standard reference book  \cite{Kamke} if $\kappa(r)$  is continuous
and preserves its sign inside an interval ($r_1,r_2$), and the potential $V(r)$ is
smooth enough, equation (\ref{67}) creates a self-ajoint eigenvalue problem, provided that
the boundary conditions at $r=r_1,r_2$   belong to the so-called
regular class. Then a complete orthonormal set of
 functions is associated to the problem (\ref{67}), the function $\kappa (r)\rmd r$ acting
as a measure in the orthonormality relations and the generalized Fourier expansions.

The zero boundary conditions at the box walls $(r_1=r_0\exp (-L/r_0)$, $r_2=R)$ proposed above
in this subsection, are regular, besides the measure (\ref{68}) has in sector IV a definite sign.
So, the conditions for the complete set to exist are fulfilled. We are going to explicitely
accomplish their construction, as well as build the matrix Green functions and the set of
Lippmann-Schwinger equations in the forthcoming paper.
\section{Concluding remarks}
\label{Conclusion}
We presented a new physical approach to the singular potential problem, using the
$1/r^2$  attractive potential as an example. A great
deal of symmetricity in handling the two singular points $r=0$ and $r=\infty$ of the
radial  \Sch is peculiar for this approach. The singular centre is analogous to
the remote infinity in the sense that
it also produces incoming and outgoing waves to form the scattering process. The waves,
incoming from and outgoing to the centre contribute into the unitarity in the same way
as the ones incoming from and outgoing to infinity. The interaction at short distances
disappears, the same as it does at large distances. This was seen after a transformation
 of the \Sch (\ref{22}), (\ref{23}), that maps the origin to minus infinity, had been
performed. This transformation turns the initial singularity into a growing potential,
that locks the particle away from the outer world, but leaves it free in the origin and
supplies it with a sufficiently ample volume element - a singular measure $\rmd r/r^2$ -
to be living in.
The strength of the singular potential Im$\la$ becomes a momentum-like quantum number
in the inner world. As long as the above confinement regime is maintained, the centre absorbs
all what it emits and hence remains stable. On the contrary, there is a two-channel regime,
where  exchanges between the inner and outer worlds take place. There a deconfinement -
we called it
escaping to infinity in the body of the paper - may occur, as well as its inverse - the
absorption. We believe that this unstable regime may be responsible for transitional processes
in the course of life of a system, described by the singular potential. During the deconfinement/absorption
the particle changes its quantum number following the conservation law (\ref{64}): that what was
the momentum in the outer world becomes the strength-of-the-singular-potential quantum number
in the inner. The overall measure (\ref{66}) serving the two-channel regime corresponds to two
different volume elements (metrics) - its limits into the inner and outer worlds.
 An arbitrary dimensional parameter $r_0$, introduced by the transformation
(\ref{22}), (\ref{23}) or, alternatively, by (\ref{64}) viewed upon as reparametrization of
states, is intrinsic in the system. It characterizes the size of the system, or the spread border
between the outer and inner worlds.

The above pattern contains many features, customary in quantum chromodynamics, such as
asymptotic freedom and confinement. It is notable that these phenomena are reproduced basing on
a very simple and also very fundamental model, proposed by nonrelativistic quantum mechanics,
which makes the primary basis of elementary particle theory.
The singular case considered also reveals many features of a black hole beyond GRT. To use a
black hole as an elementary particle prototype is sometimes considered a tempting possibility.
We believe that the features under discussion should be laid into the foundation of QFT
at the level of first and second quantization and might also modify its basic axioms, especially
in what concerns the completeness of asymptotic states.

\ack The author is idebted to Professor B.L.Voronov for helpful discussion.

\end{document}